\documentclass[10pt]{article}

\usepackage{amsmath}
\usepackage{amssymb}

\usepackage{graphicx}

\usepackage{cite}

\usepackage{color} 

\usepackage[labelfont=bf,labelsep=period,justification=raggedright]{caption}

\makeatletter
\renewcommand{\@biblabel}[1]{\quad#1.}
\makeatother

\date{}

\begin{document}

\begin{flushleft}
{\Large
\textbf{Emergent Properties of Tumor Microenvironment in a
Real-life Model of Multicell Tumor Spheroids}
}
\\
Edoardo Milotti $^{1,2,\ast}$ and
Roberto Chignola $^{2,3}$\\
\bf{1} Dipartimento di Fisica, Universit\`a di Trieste, Trieste, Italy\\
\bf{2} I.N.F.N.-Sezione di Trieste, Trieste, Italy\\
\bf{3} Dipartimento di Biotecnologie, Universit\`a di Verona,  Verona, Italy\\
$\ast$ E-mail: edoardo.milotti@ts.infn.it\\
\end{flushleft}

\markboth{Model of Tumor Spheroids}{Model of Tumor Spheroids}

\section*{Abstract}
{
Multicellular tumor spheroids are an important {\it in vitro} model of the pre-vascular phase of solid tumors, for sizes well below the diagnostic limit: therefore a biophysical model of spheroids has the ability to shed light on the internal workings and organization of tumors at a critical phase of their development. To this end, we have developed a computer program that integrates the behavior of individual cells and their interactions with other cells and the surrounding environment. It is based on a quantitative description of metabolism, growth, proliferation and death of single tumor cells, and on equations that model biochemical and mechanical cell-cell and cell-environment interactions. The program reproduces existing experimental data on spheroids, and yields unique views of their microenvironment. Simulations show complex internal flows and motions of nutrients, metabolites and cells, that are otherwise unobservable with current experimental techniques, and give novel clues on tumor development and strong hints for future therapies.}

\section*{Author Summary}
{
When dealing with tumors, a strong emphasis is usually placed on the detailed molecular machinery of individual cells. However, their interactions with the environment and their collective behavior are equally important, and largely unexplored. Biologists study these interactions in the laboratory, in cultures of small spherical clusters of cells called ``tumor spheroids''. Unfortunately it is very difficult to exploit the full potential of these experimental studies and extract the precious informations withheld in the core of spheroids. We have developed a computer model that helps understand the interaction between cells and their environment, and establishes a bridge between the microscopic world of molecular interactions and the macroscopic properties of spheroids. This computer model is a sort of laboratory where it is possible to perform virtual experiments on tumor spheroids and their environment, and gain unhampered access to all useful informations on the internal workings of these model tumors. Eventually, we are rewarded with novel and unexpected views of the microstructure of tumor spheroids.
}

\section*{Introduction}
Multicellular tumor spheroids (MTS) stand out as the most important {\it in vitro} model of pre-vascular solid tumors \cite{r3,r4,r5,r6,Gott2006,Fried2007,LinChang2008,Hirsch2010}. MTS often have a regular, almost spherical structure, and their apparent simplicity has led to repeated attempts to capture their features with neat mathematical models. However, the absence of vascularization and the near sphericity hide an internal complexity which is not easy to tame either with analytic mathematical models \cite{r7,r8,r9,r10}, or with numerical models based on rough simplifications of the biological settings such as cellular automata or other lattice-based models \cite{r11,r12,r13,r14}. Moreover the presence of a growing necrotic core \cite{r3} and of an extracellular matrix \cite{r15}, the appearance of convective cell motions \cite{r16}, and the heterogeneous response to chemotherapics \cite{r17}, point to the importance of MTS as an in vitro model of tumors, and most of all to their relevance to understand tumor heterogeneity, but they also point to the difficulties of producing a useful, predictive model of MTS. 

The appearance of widely different resistance phenomena to antitumor therapies in similarly grown, isolated MTS of the same cell type \cite{r17} indicates that random fluctuation phenomena play an all-important role in the growth kinetics of MTS. It is well-known that the discrete events at the single-cell level (like transitions from one cell-cycle phase to the next, mitosis, cell death, etc.) do display some randomness, and one can pinpoint the source of large-scale variability on these fluctuations, as they are amplified and propagated by cell-cell and cell-environment interactions. Thus, the complexity of MTS development can only be captured by a fine-grained, multiscale model, and we need a mathematical description at the single-cell level. Since cells communicate with other cells and the environment, the other actors of this complex play are the concentration gradients of important molecular species that depend on the structure of the extracellular space and of the facilitated transport processes into and out of individual cells, and the mechanical forces that push and pull cells as they proliferate with repeated mitoses and then shrink after death \cite{r18}. These processes mix with complex nonlinear interactions between the biochemical and the mechanical part, and this highlights again the importance of an effective model at the single-cell level.

On the basis of such motivations, we have developed a numerical model of MTS that incorporates a working model of single cells \cite{r19,r20}. We have first put forward a broad outline of its structure in reference \cite{r21}, and it differs from other models developed in the past \cite{r7,r8,r9,r10,r11,r12,r13,r14} because it captures at the same time both the basic features of cell metabolism, growth, proliferation and death, and provides a true lattice-free calculation of cell motions, as they are pushed and pulled by the forces exerted by dividing cells, by the growth of other cells, and by the shrinking of dead cells. We also wish to stress that the model parameters are either derived from experiment or are deduced from reasonable theoretical arguments, so that, essentially, there are no free parameters -- there can only be some residual variability in biophysically meaningful ranges -- the model is truly predictive, and the results are not merely qualitative but quantitative as well. 

Here we illustrate in broad terms the structure of the program and report the results of the first simulations of single spheroids (technical implementation details are relegated to the supporting text). We find that the simulations agree quite well with experimental measurements on real spheroids, and show unexpected and important internal patterns. Moreover, we wish to stress that the methods delineated in this paper represent very general practical solutions to problems that are common to any simulation of cell clusters, and they are just as important.

\section*{Biochemical behavior of individual cells}

\begin{figure}[!ht]
\begin{center}
\includegraphics[width=5in]{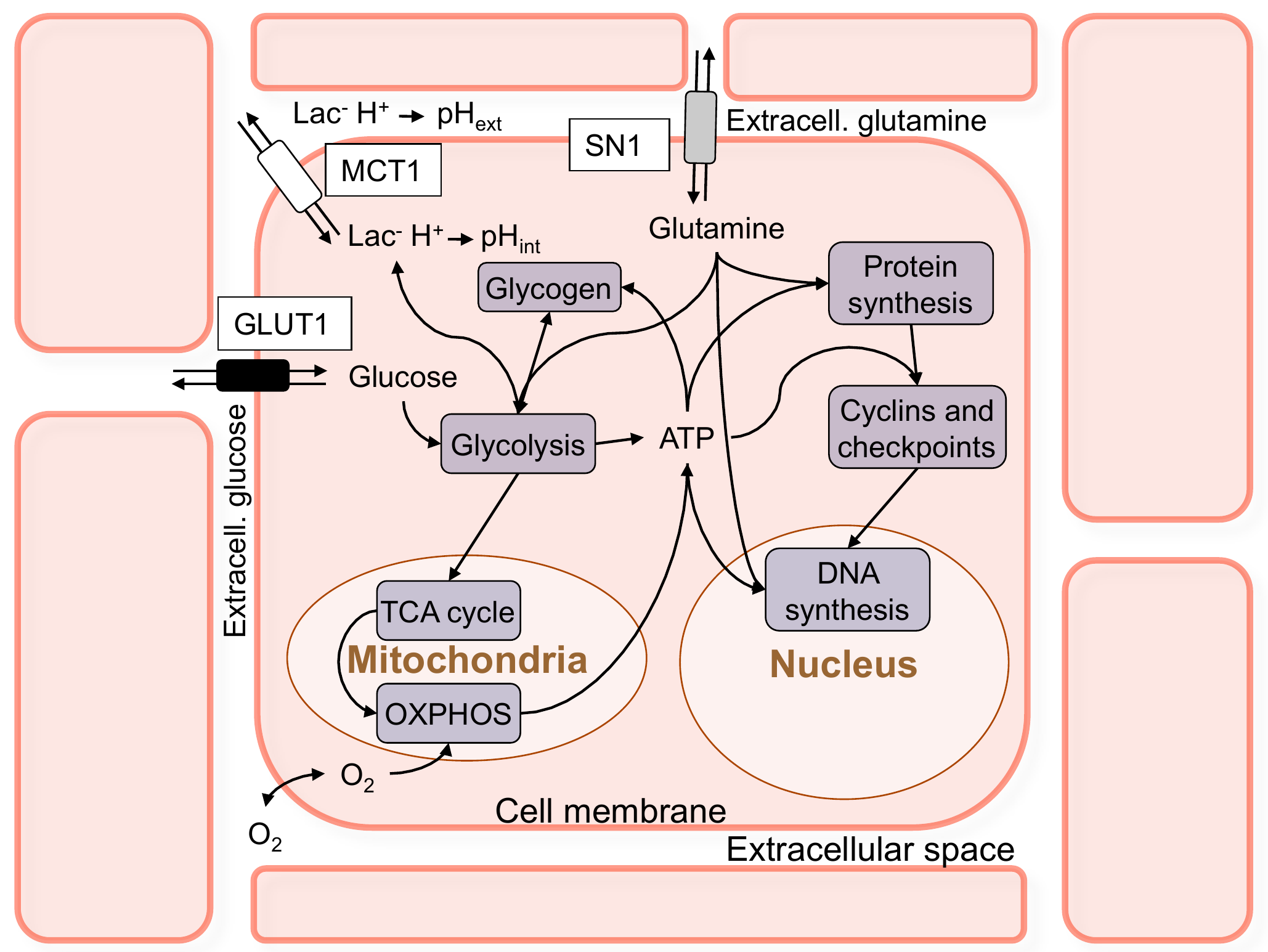}
\end{center}
\caption{
{\bf  Rough sketch of the biochemical pathways incorporated in the model of single cells.} We take into account the main metabolic pathways (glycolysis, oxidative phosphorylation through the TCA cycle and gluconeogenesis), including the role of mitochondria in the production of ATP. The model also includes protein and DNA synthesis, and checkpoints controlled by representative members of the cyclin family. The single-cell model has two spatial compartments (the inside of the cell and its immediate neighborhood, the extracellular space that surrounds it) and transport of substances between these compartments is regulated by transporters on the cell membrane that are also included in the model. The extracellular space of each cell communicates by simple diffusion with the neighboring extracellular spaces and with the environment. The complete map of the biochemical pathways is shown in figure SF2 in the supporting information text.}
\label{fig1}
\end{figure}

The elementary building blocks in this model of MTS are the individual tumor cells that behave as partly stochastic automata \cite{r19,r20}. Figure \ref{fig1} summarizes the biochemical pathways that are included in the single-cell model: cell metabolism is driven by oxygen, glucose and glutamine, and transforms these substances into energy molecules, molecular building blocks and waste products, following the well-known biochemical reaction chains \cite{r22}. Further details can be found in the original papers \cite{r19,r20} and in the supporting text, which also includes important upgrades to the original model \cite{r19,r20}.

In the present version of the program, the stochasticity is mostly concentrated in the discrete events: for instance, mitochondria are partitioned at random between daughter cells at mitosis, and cells can die because of metabolite accretion, according to a Poissonian cytotoxicity model (see the supporting text).

We remark that in this approach glutamine also stands for the wider class of aminoacids, and lactate is the paradigm of all metabolites: we use the concentrations of glutamine and lactate to represent these two classes of substances in phenomenological parameterizations wherever needed. Similarly we use the number of mitochondria and ATP content to model the dynamics of cell volume; the single-cell model also contains representative members of the cyclin protein class to compute the passage of checkpoints and entry into the different cell phases \cite{r19,r20,r23,r24}, and finally into mitosis (see also figure SF1 in the supporting information for a sketch of the cell cycle in the simulation program).

The complete map of the biochemical pathways included in the simulation program is shown in figure SF2 in the supporting text. This map comprises only the most basic pathways, however we cannot afford to introduce a more complex network at this stage of program development. Indeed, our final aim is the simulation of MTS with a volume as large as 1 mm$^3$, which corresponds to more than one million cells, so that simulation results overlap actual experimental measurements \cite{r17,r25,r26}. Since the differential system involves 19 independent biochemical variables per cell, we must eventually integrate at least 19 million coupled nonlinear differential equations for the biochemical cell variables alone (this grows to at least 25 million equations when we include the position and velocity variables), and thus even this minimal single-cell model leads to a daunting computational task (see the supporting text for further details on the algorithmic complexity of the program).

\section*{Reaction-diffusion processes and the environment}

Substances like oxygen are transported into and out of cells by normal diffusion while molecules like glucose require facilitated diffusion processes. This means that cell membranes play an important role for substances like glucose, and that in this case the diffusion of each such molecular species towards cells in the inside of a spheroid needs the free volume in the extracellular space to proceed, and that we must model this space as well as the cells to obtain a realistic simulation. We have shown how to do this in reference \cite{r27}, where we have also discussed ways to tame the exceedingly high stiffness of the very large set of reaction-diffusion and transport equations that arise in this context (see also the supporting text). The external environment itself is included in these equations, and evolves in time as well. In the present model, each cell contributes 15 internal variables and 4 extracellular variables: these extracellular variables are the masses of oxygen, glucose, glutamine and lactate in the extracellular volume surrounding the cell. Because of its smallness, the extracellular space has an extremely short characteristic filling time, which can be as fast as few tens of microseconds. On the other hand, the macroscopic features of MTS evolve over times as long as months (i.e., times of the order of $10^7 \mathrm{s}$), and thus the numerical integrator must be able to handle phenomena that span 12 orders of magnitude in time \cite{r27}. The internal biochemical reactions included in the numerical model are much slower and their fastest characteristic times are only as low as $0.1 \mathrm{s}$, much longer than the diffusion times \cite{r27,r28}.
The topology of diffusion in the extracellular spaces is obviously dictated by the cells themselves, and the program uses the network of cellsÕ centers as the scaffolding for the corresponding discretized diffusion problem. The links between the cells' centers -- i.e., the proximity relations -- are provided by a Delaunay triangulation \cite{r29,r30}, which is computed repeatedly \cite{r31} as the cluster of cells grows and rearranges itself under the pushes and pulls of volume growth, mitosis, and the shrinking of dead cells (see also figure SF3 in the supporting information). Moreover, the proliferation of cells means that both the number of cells and the total number of links steadily grow, and that the differential system of equations that model metabolism, transport and diffusion changes all the time, and becomes increasingly complex. The 3D Delaunay triangulation itself is not an exceedingly heavy computational burden for the program, as it turns out that efficient algorithms can compute it, on average, with $O(N)$ time computational complexity \cite{r31,r32,r33}, so that this algorithm is indeed feasible for very large clusters of cells.

\section*{Biomechanical evolution of the simulated MTS}

Real cells have passive viscoelastic mechanical features, but they also move actively under the pushes of their own cytoskeleton, and to the best of our knowledge there is no comprehensive model of cellular biomechanics \cite{r36,r37}. Thus, we resort once again to phenomenological simplifications, and the first and foremost is that our cells are stretchable spheres, characterized by their radius, and a few other parameters that specify their viscoelastic properties (see the supporting text for a more detailed description and the list of parameters). We also specify a pairwise interaction force between cells, repulsive when a cell pushes against a neighbor, and attractive when we try to detach it from its neighbor. For small deviations from the equilibrium distance, we assume that the interaction force is described by the Hertz model (explained in the  supporting text), while for large deformations due to compression we set the force to a fixed saturation value, and for large distances the attractive force decays to zero (see figure SF4 in the supporting information). The description of the interaction forces is tuned to hold also during mitosis (see the supporting text and figure SF5). Even though this is a rough approximation of the overall mechanical behavior of cells, there are many details that must be managed to make it work, and they are all described in the supporting text.

Here the Delaunay triangulation that we use as the scaffolding for the diffusion problem turns out to be useful once again: the same cell-cell links also define the set of neighbors of each cell, and therefore the global problem of computing the pairwise interactions between cells can be reduced to a single loop over all cells and the small limited number of their immediate neighbors, so that this operation has an $O(N)$ computational complexity only -- and it does not grow when we include the cost of the Delaunay triangulation \cite{r33} -- instead of the $O(N^2)$ complexity of generic pairwise interactions.

\section*{Results}

\begin{figure}[!ht]
\begin{center}
\includegraphics[width=6in]{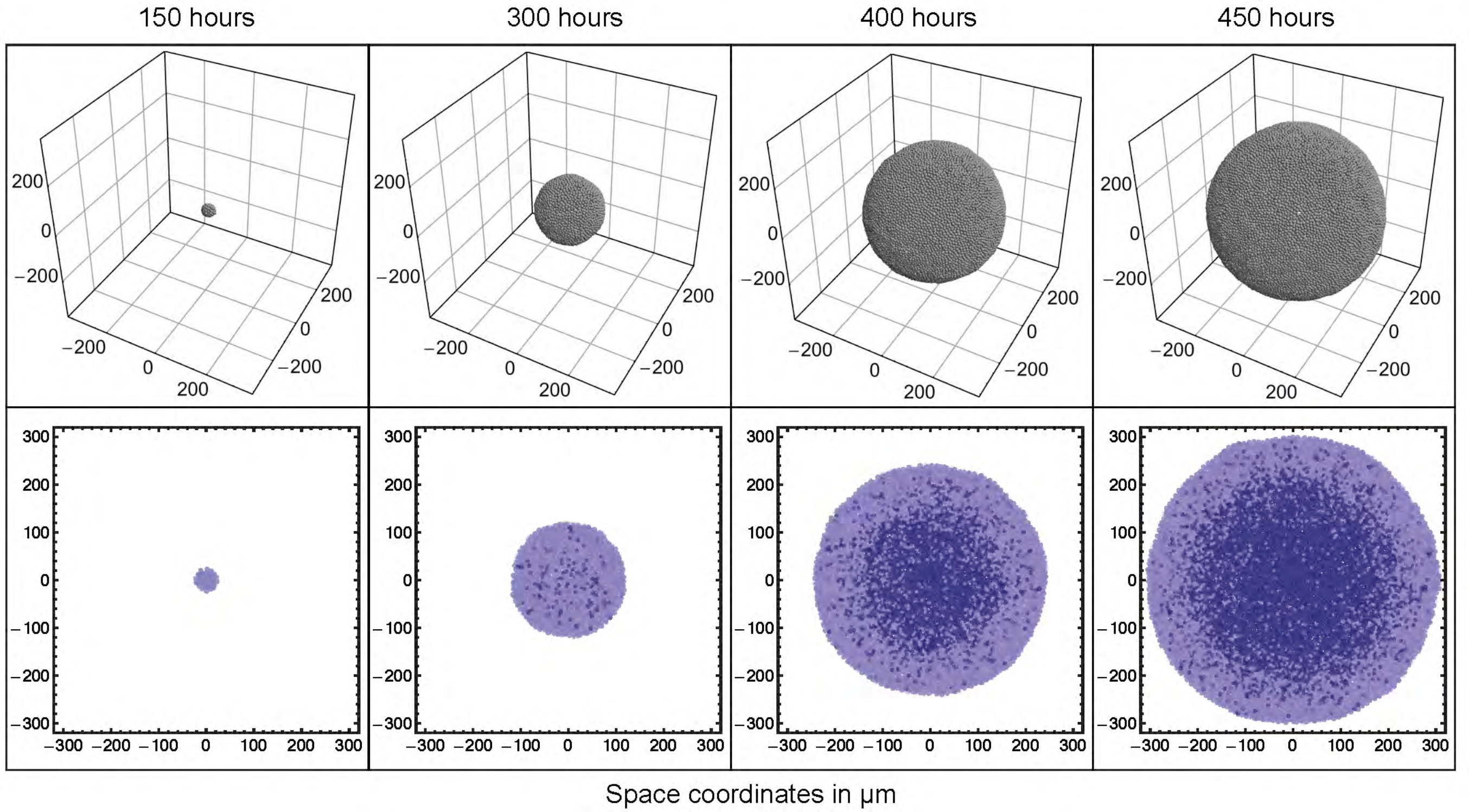}
\end{center}
\caption{
{\bf  Snapshots of one simulated spheroid taken at different times.} As the spheroid grows, a necrotic core develops in its central region, just as it happens in real spheroids. The size of the necrotic core and of the viable cell rim match real measurements.}
\label{fig2a}
\end{figure}

\begin{figure}[!ht]
\begin{center}
\includegraphics[width=1.5in]{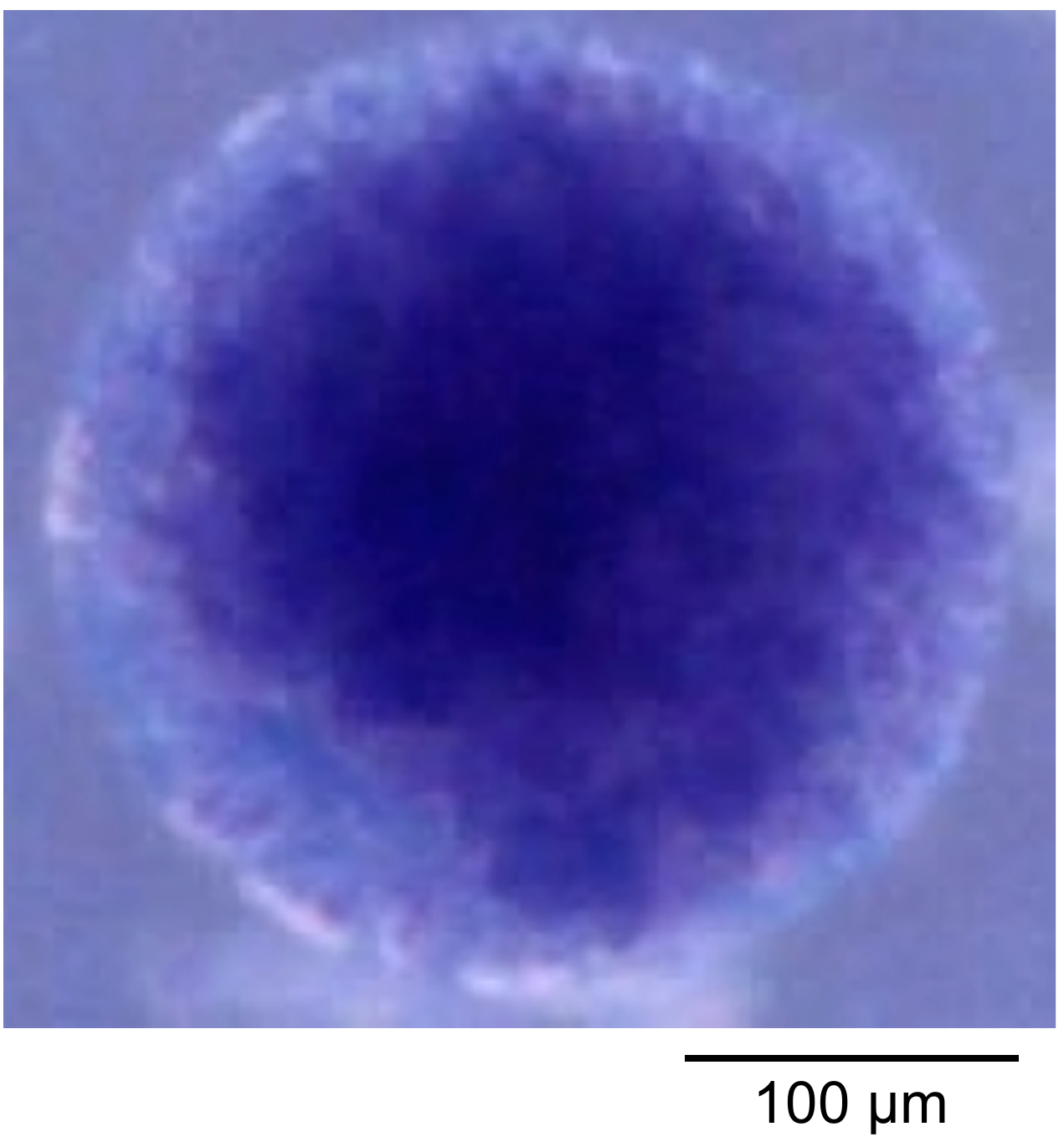}
\end{center}
\caption{
{\bf  Photograph of a spheroid grown in vitro from HeLa cells in agar.} The spheroid is colored with trypan blue to mark dead cells, where the necrotic core is clearly visible. The agar contains the spheroid and helps in obtaining a better spherical shape with HeLa cells, but also stifles spheroid growth because it reduces the effective diffusion coefficients in the nourishing medium, so that it cannot be directly compared to the simulated spheroid in the second column of figure \ref{fig2a} (which has the same size), while it is similar to the larger spheroid in third column.}
\label{fig2b}
\end{figure}

\begin{figure}[!ht]
\begin{center}
\includegraphics[width=6in]{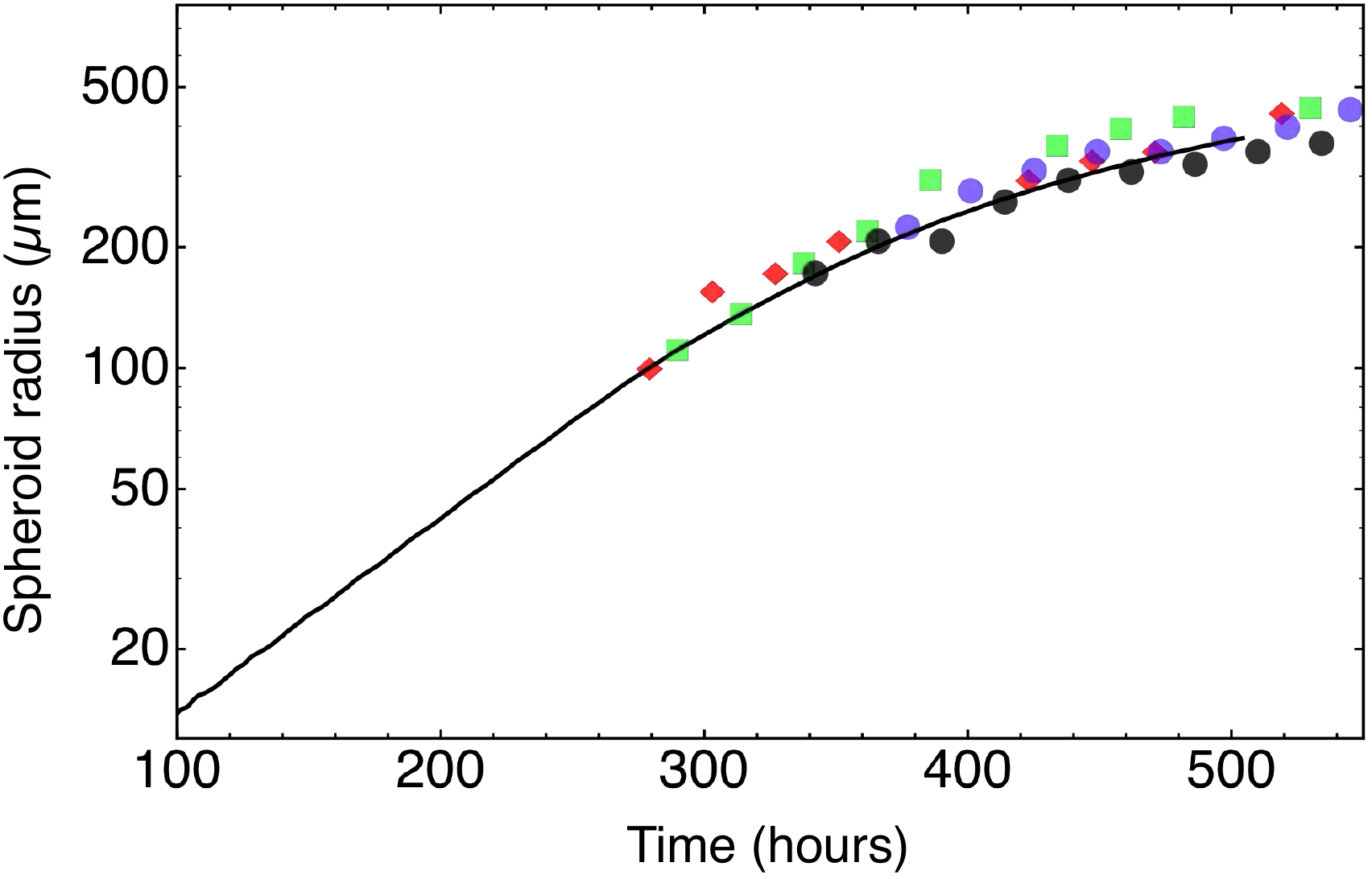}
\end{center}
\caption{
{\bf  Growth curve of a simulated tumor spheroid (solid line).} The run parameters used in this case are listed in tables ST3, ST4 and ST5 in the supporting information text. The symbols denote data points taken in different in vitro experiments: squares = FSA cells (methylcholantrene-transformed mouse fibroblasts) \cite{r43}; diamonds = MCF7 cells (human breast carcinoma) \cite{r17}; circles = 9L cells (rat glioblastoma) \cite{r25}}
\label{fig3}
\end{figure}

The first and most obvious result is the outstanding match of the growth curves of simulated spheroids with those of real spheroids: figure \ref{fig2a} shows a few stages of the growth of a simulated spheroid (a real spheroid is shown for comparison in figure \ref{fig2b}), while figure \ref{fig3} compares the growth curve of a single simulated spheroid with the growth curves of real spheroids grown {\it in vitro}. Here we see that the growth curves are very much alike, and we found that simulation runs with different parameters -- in the biophysically meaningful ranges -- produce very similar growth curves, in spite of structural internal changes: the growth curves are thus rather robust with respect to parameter changes. 

\begin{figure}[!ht]
\begin{center}
\includegraphics[width=5in]{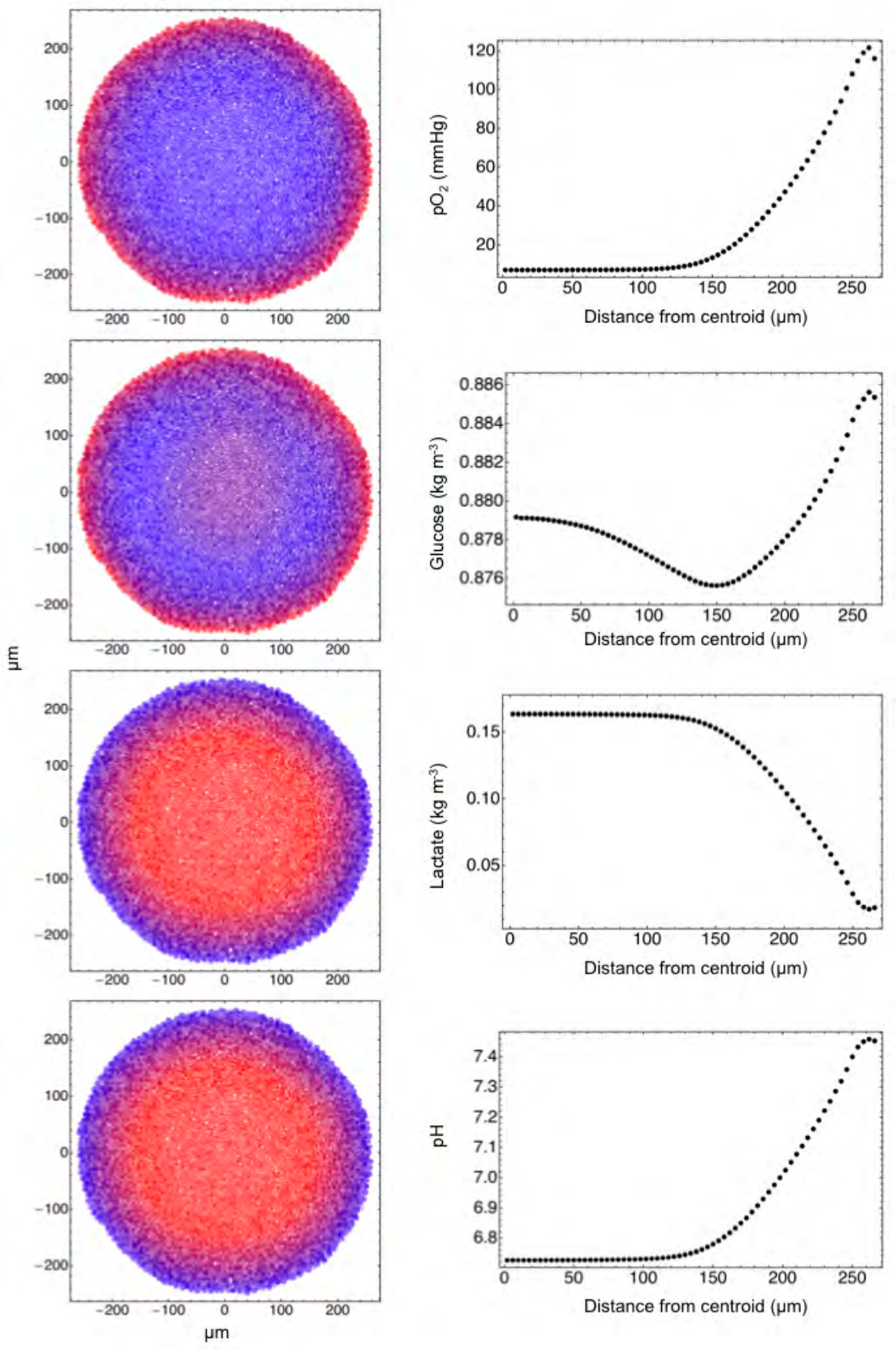}
\end{center}
\caption{
{\bf  Concentrations in the simulated spheroid.} The color coded figures on the left show the partial pressure of oxygen, the concentrations of glucose and lactate in the extracellular spaces, and the pH of the extracellular environment (high values = red, low values = blue). The corresponding plots in the right column show the average values of these quantities vs. the distance from the centroid of the tumor spheroid. The small oscillations in the plots close to the spheroid surface are due to fluctuations in the averaging procedure, because the spheroid is slightly nonspherical. }
\label{fig4a}
\end{figure}

\begin{figure}[!ht]
\begin{center}
\includegraphics[width=6in]{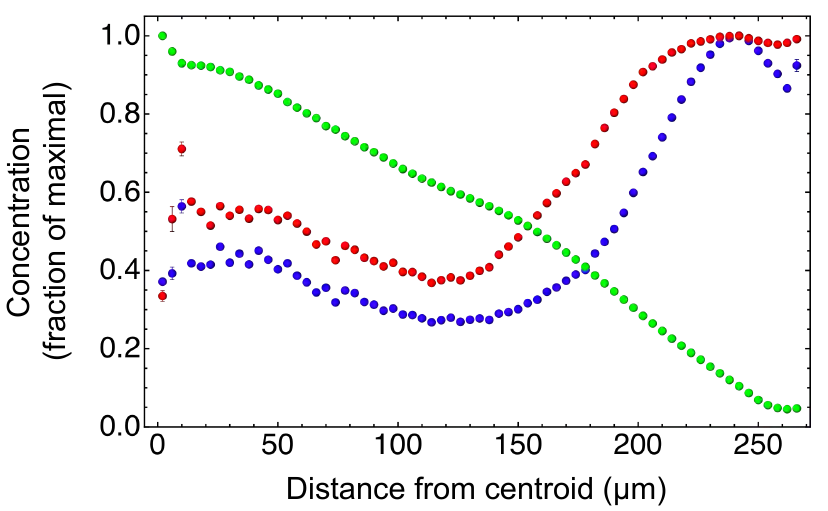}
\end{center}
\caption{
{\bf   Plots of the normalized average intracellular concentration of lactate (green), glucose (blue), and ATP (red).} These plots have been obtained in the same simulation and at the same time step as the plots of figure \ref{fig4a}, and each concentration is normalized to its peak value. These plots indicate that cell death in the central region is due both to the accumulation of metabolites (lactate) and to metabolic stress (starvation).}
\label{fig4b}
\end{figure}

Several experiments \cite{r37,r38,r39,r40,r41,r42} have yielded many accurate measurements of oxygen and glucose concentrations and other quantities vs. spheroid radius; these values are part of the output of our simulation program as well (see figure \ref{fig4a} and figure \ref{fig4b}), and a comparison with the experimental data is shown in the table. On the whole the agreement of simulation data of single spheroids with the experimental values is quite good, and we wish to stress that this is not the result of a fit {\it a posteriori}, but rather of the {\it a priori} choice of model features and parameters. These results qualify as true predictions of the numerical model. 

\begin{figure}[!ht]
\begin{center}
\includegraphics[width=6in]{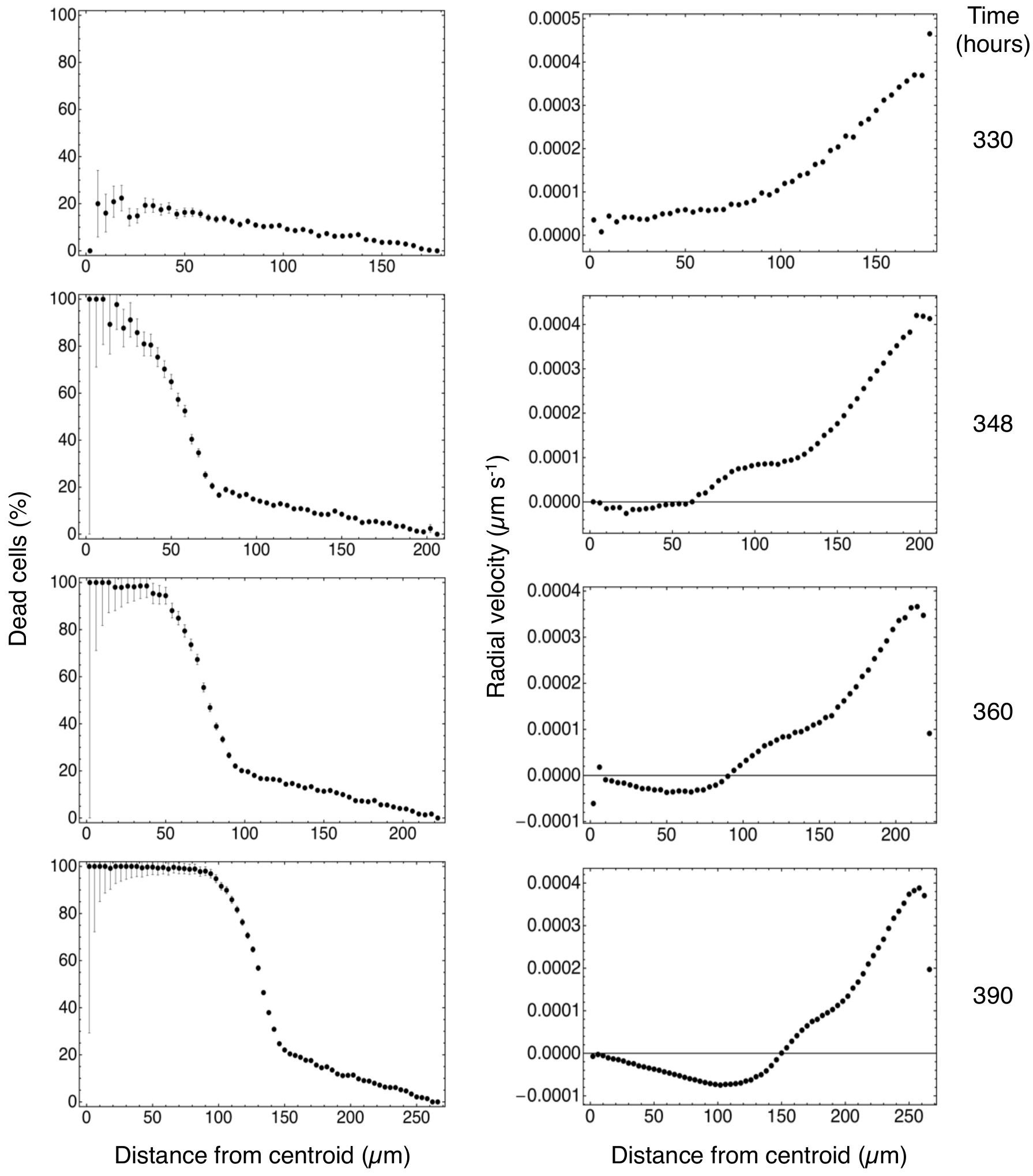}
\end{center}
\caption{
{\bf Fraction of dead cells (left column) and average radial velocity (right column) at different times.} As the spheroid grows, the necrotic core becomes increasingly well defined, and as dead cells shrink, the radial velocity changes sign and a marked inward motion characterizes the central region.}
\label{fig5}
\end{figure}

The necrotic core of spheroids is another important feature that is well reproduced in the simulations, and it is clearly visible in central slices of the simulated spheroid in figure \ref{fig2a}. The simulations also provide detailed, quantitative snapshots of the necrotic core dynamics; the left column of figure \ref{fig5} shows the percentage of dead cells vs. distance from the centroid of a simulated spheroid at different times. In these snapshots we can clearly observe the formation of the sharp step that marks the edge of the necrotic core.

\begin{figure}[!ht]
\begin{center}
\includegraphics[width=6in]{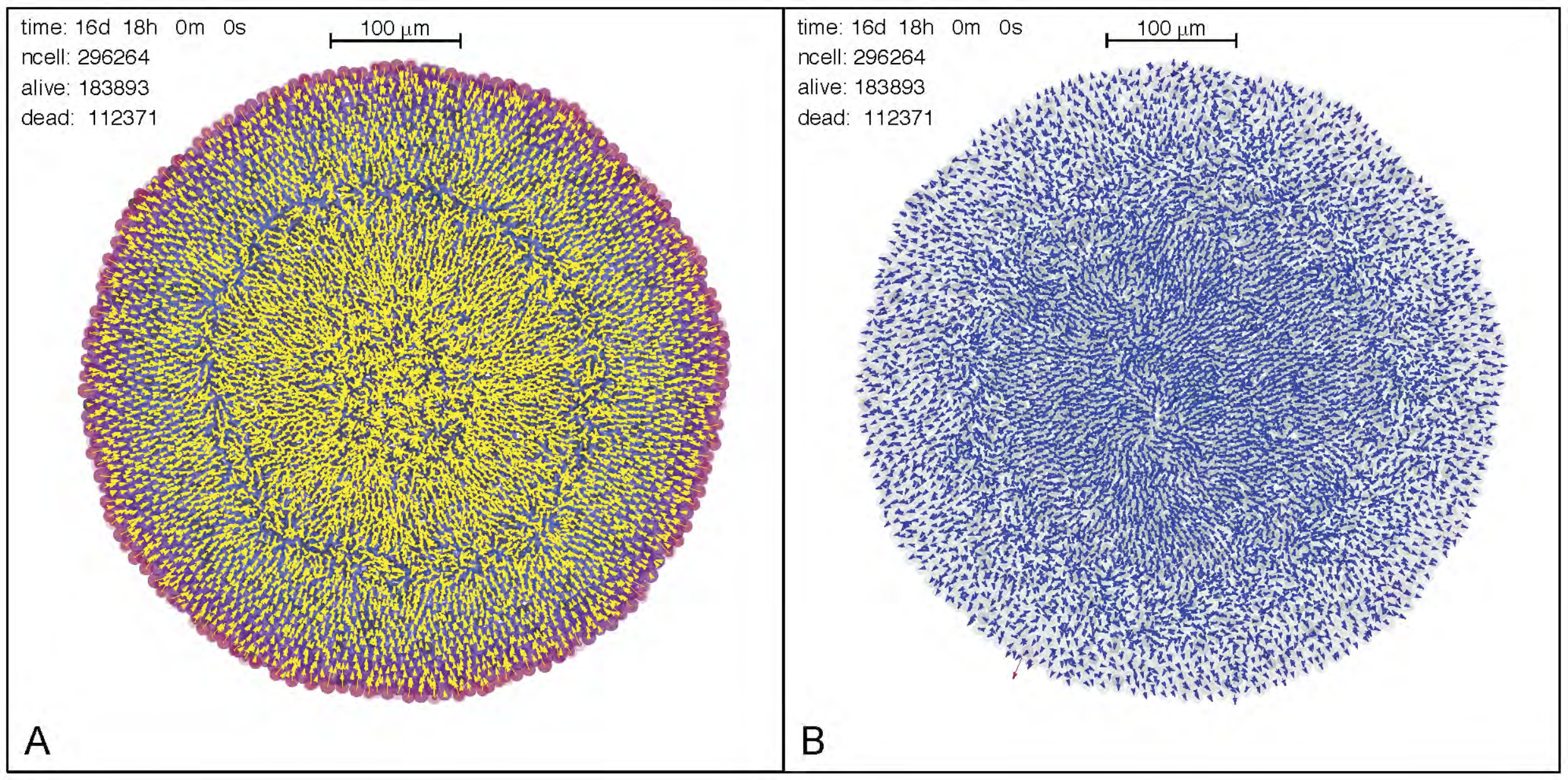}
\end{center}
\caption{
{\bf Two views of the microstructure of a simulated spheroid, with about $\mathbf{500 \mathrm{\mu m}}$ diameter and 296264 cells (183893 live cells + 112371 dead cells).} (Left panel): flow of extracellular glucose along a central section of the tumor spheroid (yellow arrows) superposed on the plot of glucose concentration. The length of the arrows is proportional to the glucose flow intensity projected on the plane of the section. At this stage, the necrotic core is contracting as dead cells gradually shrink, and this leads to a slow outward flow of the glucose stored in the extracellular spaces in this central region. We observe that such a behavior depends on the effective diffusion coefficient of glucose, and it disappears completely when the diffusion coefficient is high enough. This also suggests that the flow of glucose and other substances, like therapeutic drugs, is strongly dependent on the biochemistry and structure of extracellular spaces, and even small changes can lead to markedly different internal spheroid morphologies. (Right panel): individual cell velocities in the simulated spheroid. This is the same central section as in the left panel, and the velocity vectors are projected on the plane of the section. The length of each vector is proportional to the projected speed. The velocities in the viable rim show a coherent outward motion, while the velocities in the necrotic core show a rather orderly inward motion, with some vortices due to local residual cell proliferation. The region in-between is somewhat chaotic and the global structure of this plot mirrors that of the glucose flow shown on the left. The supporting information includes higher-quality versions of these figures and those of other flows.}
\label{fig6}
\end{figure}

These results indicate that the simulation program is reliable and robust and reproduces -- both quantitatively and qualitatively -- known experimental results. However, it yields much more than just successful comparisons: figure \ref{fig6} shows two views of the spheroid microenvironment that at present would be unobtainable by other means at this level of resolution. The left panel of figure \ref{fig6} is a plot of the flow of glucose in the extracellular spaces of a mature spheroid, superposed on a density plot of extracellular glucose concentration, and it shows -- rather unexpectedly -- that there is an outward flow of extracellular glucose from the central necrotic region. In the external, viable rim the flow is inward bound, and there is a spherical shell where the flow is stationary. The right panel of figure \ref{fig6} shows the corresponding plot of cell velocities in the same central slice, and we see that the velocity vectors point outward in the viable rim, while there are well-formed vortices in the central region, and the region in-between displays distinctive chaotic motions: these three regions closely match the three regions in the left panel. The right column in figure \ref{fig5} shows radial velocity vs. distance from the centroid of the simulated spheroid, and sheds some more light on the nature of this structure: as more and more cells die and the necrotic core forms, the dead cells shrink and the core contracts. The contraction of the necrotic core expels the residual glucose in the extracellular spaces and produces the observed outward flow. We found that this behavior is strongly dependent on the particular value of the diffusion coefficient and on the metabolic activity of cells. In some simulations -- where we used a lower value for the effective diffusion coefficient of oxygen -- we observed a similar structure with oxygen as well. We remark that in the case of lactate we found no such structure, and we obtained a pH value -- derived from the distribution of lactate inside the spheroid --  that is very close to experimental measurements: this indicates that the discretized reaction-diffusion scheme used in the simulation program performs correctly, and that the observed flows are not algorithmic artifacts.



\section*{Discussion}

Although the program described in this paper is based on a model of individual cells that includes only the basic cell functions, the simulation results compare very well with experimental measurements, and give strong hints on the sources of individual spheroid variability. Moreover, the images obtained in single runs reveal unexpected and interesting correlations and an elaborate structure of the tumor microenvironment that could never be observed before. 
This unexpected, complex microstructure -- the formation of different regions, and the flows that characterize them, along with the complex velocity field --  can be discerned in the flows of the other substances, though not all of them, according to their effective diffusion coefficient and their metabolism: the figures of these flows are shown at full-size as supporting information. Thus if we suppose that, in a more complete description, there are $N$ substances that characterize the spheroid microenvironment, and assume that the spherical shell that divides the two main regions lies in the same position for all of these substances and that their effective diffusion coefficients are uncorrelated, then $2^N$ different spheroid structures are determined by diffusion alone. The variation of some critical parameter (e.g., a slight change in the metabolic activity due to local fluctuations in the number of dead cells, and thus a change in the effective diffusion coefficients) can potentially act as a switch and determine widely different fates for similar spheroids. This variability cannot be discerned from growth experiments: the simulations that we have performed to date indicate that the growth curve alone is not enough to distinguish between such different states, because it does not change much even when important substances, like oxygen, diffuse in markedly different ways. These different states represent different biochemical configurations of tumor microenvironment, that might exert distinct selective pressures on cells during tumor evolution.

The spheroid microstructure that is well evidenced in figure \ref{fig6}, and in figures SF8-21 and in the movie files S1-3 in the supporting information, shows highly correlated fluctuations that produce, e.g., islets of proliferating cells in the sea of dead cells of the core, and cell and mass flows that follow preferential channels. There is a sort of spheroid-specific self-organization of the internal structure due to these correlated fluctuations. Similar cell flows have been observed in the lab and a recent review has stressed the great significance of such findings \cite{Deis}: the simulations suggest that the whole topic of cell flows and extracellular diffusion should be investigated further.  On the basis of the simulation results, we also conjecture that the flow of therapeutic drugs may be diverted as well, and let some viable, proliferating tumor cells escape treatment. This means that the simulation program could eventually become an important tool to design novel treatment schedules, and possibly validate the effects of anti-tumor drugs. 

Certainly the model is far from complete, and we plan to add soon several new features, like a basic model of intracellular acidity, now accounted for by a simple phenomenological parameterization, and the effects of pH and salt concentration on diffusion. However, already in its present form, we believe that this numerical model is a true testbed of biological complexity and a real virtual laboratory, and also a source of important biomedical clues.

\section*{Methods}

The simulation program is written in ANSI C++: this programming language was a natural choice from the very start for distinct reasons:
\begin{itemize}
\item{C++ is an object-oriented language, and in a simulation such as this, it is very natural to define objects that have a clear-cut biological meaning;}
\item{at present, C++ programming is supported by a vast array of scientific libraries, and this helps reducing program development time;}
\item{the availability of the flexible and powerful C++ library CGAL \cite{r31} that handles the computational geometry structures utilized by the program (convex hulls, Delaunay triangulations and alpha shapes);}
\item{the availability of powerful development tools and highly optimized compilers.}
\end{itemize}

\begin{figure}[!ht]
\begin{center}
\includegraphics[width=6in]{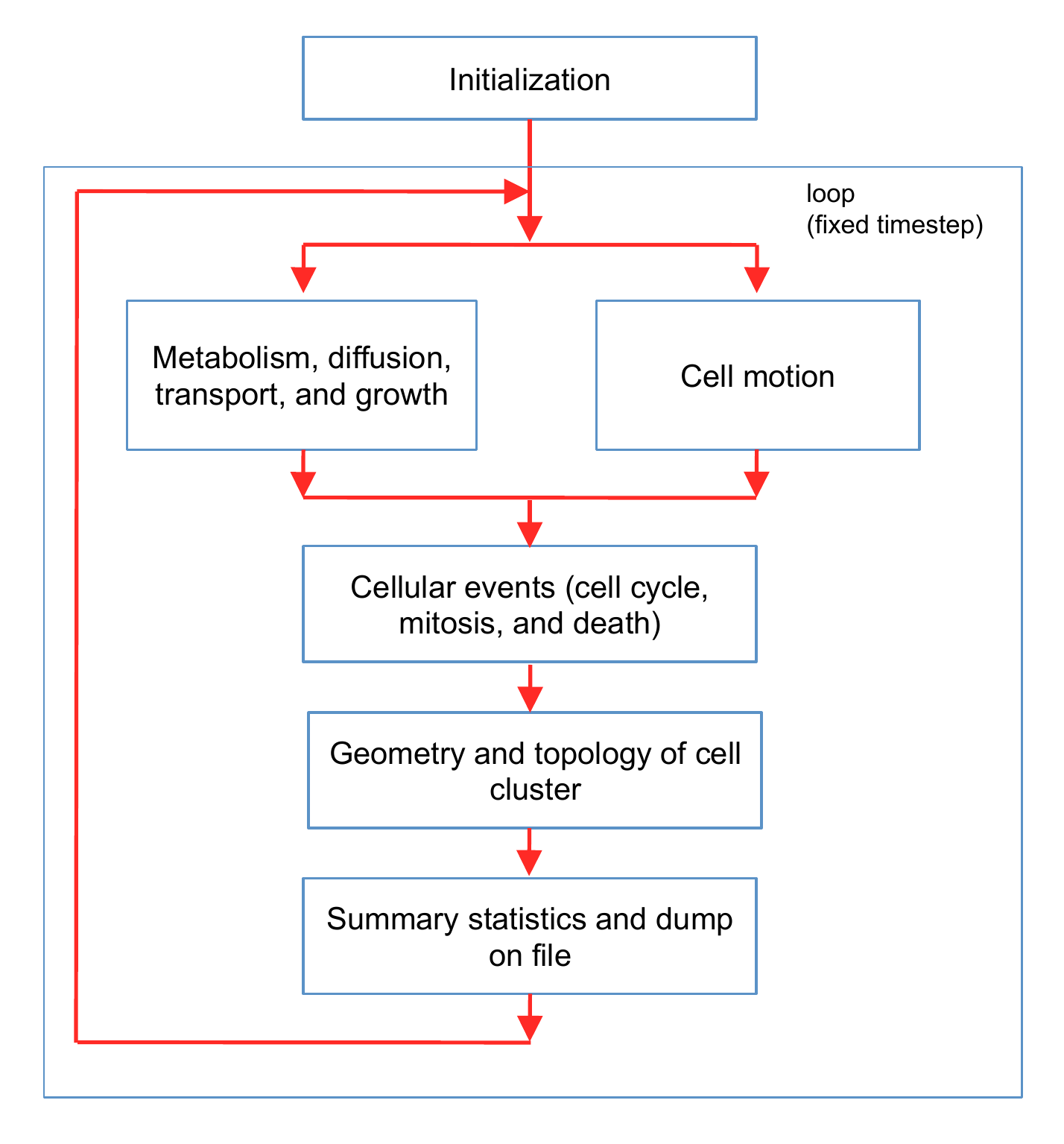}
\end{center}
\caption{
{\bf Functional blocks of the simulation program.} Program initialization is followed by a loop that performs biochemical and biomechanical calculations. This is followed by a check of the status of individual cells -- this is where we decide whether a cell advances in the cell cycle, undergoes mitosis, or dies. Next the program computes the geometry and the topology of the cell cluster, and finally it outputs intermediate statistics and results. The loop continues until a user-defined stop condition is met. Some parts of the program can proceed in parallel (like metabolism and cell motion), and we can use multithreaded code.}
\label{fig7}
\end{figure}

The structure of the program reflects the organization explained in the paper: a layout is shown in figure \ref{fig7}. The functional blocks work as follows:

\subsection*{Initialization}

At start, the internal variables of all cells are set at approximate standard values. During initialization, cells are allowed to grow and proliferate freely in an environment that is held fixed. The number of cells is also kept constant, and when a mitosis occurs one of the daughter cells is discarded. In this initial phase cells can have large oscillations of their metabolic parameters, and can occasionally step in parameter regions that would normally spell death: this does not occur here. Initialization lasts until the oscillations of metabolic parameters die out. We have determined the duration of the initialization phase observing the desynchronization of a population of initially synchronized cells: when oscillations of the relative fractions of cells in each cell-cycle phase become undetectable we estimate that cells have reached a stable state. It turns out that a simulated time of $3\cdot 10^6 \mathrm{s}$ (i.e. about 35 days of simulated time) is sufficient for initialization of cell with a period of about 20 hours. Usually the starting number of cells is quite small (normally just one cell to seed the growth of a single spheroid), and initialization executes in very short real time (a few seconds). 

\subsection*{Metabolism, diffusion, transport, and growth}

\begin{figure}[!ht]
\begin{center}
\includegraphics[width=6in]{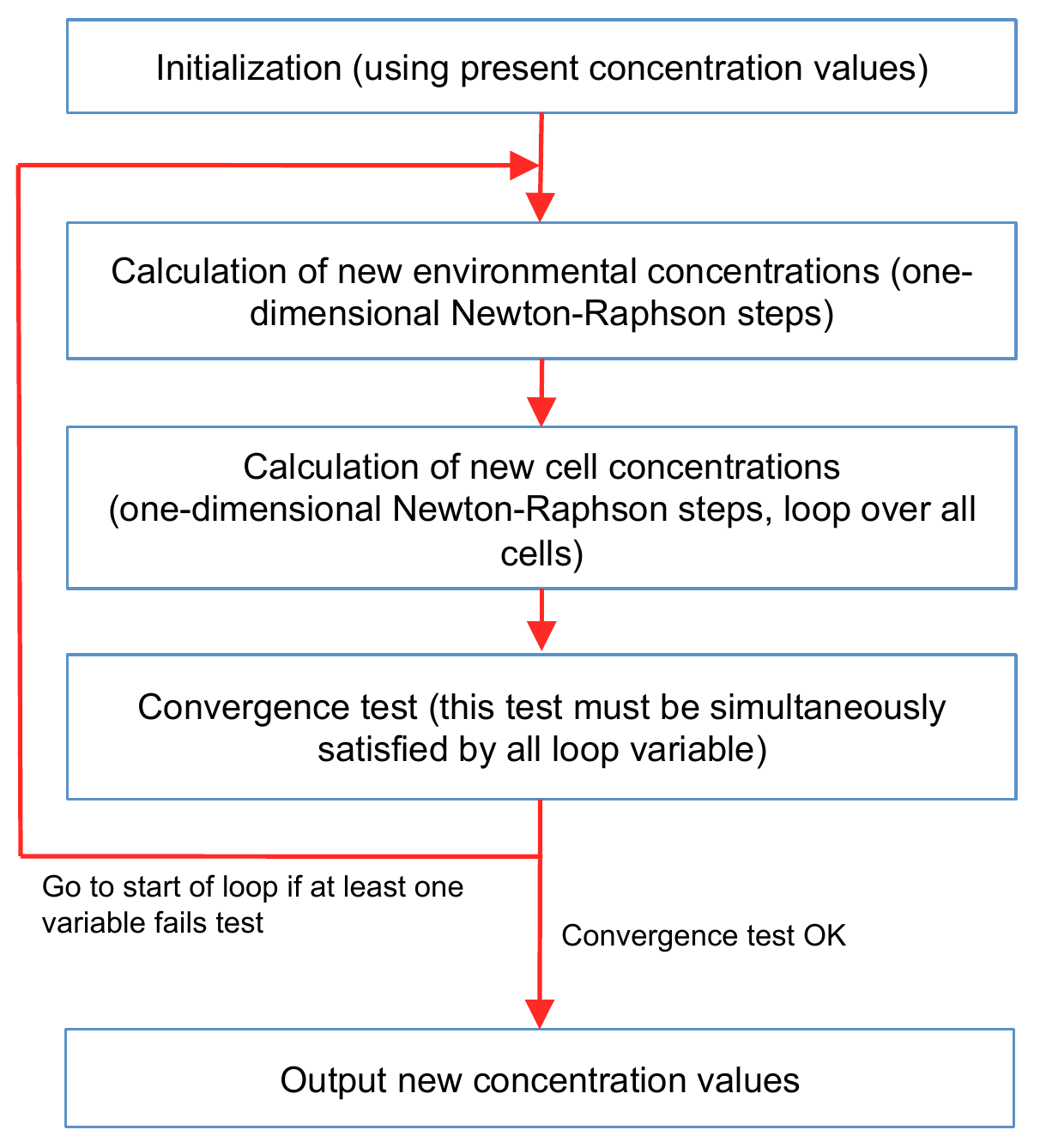}
\end{center}
\caption{
{\bf Functional blocks of the C++ method that computes metabolic and extracellular variables.} This part performs a loop that computes the solution of the nonlinear equations found in the implicit Euler integration step \cite{r27} (see also the supporting text). Although the number of variables can be quite large (more than $10^7$ variables), convergence is fast, because the initial concentration values are invariably very close to the final ones.  }
\label{fig8}
\end{figure}

This part of the program solves the combined differential system of equations that describe internal cell metabolism and diffusion in the extracellular spaces (described in detail in the supporting text), using the implicit Euler method. This leads to a system of nonlinear equations, that  are solved in turn with a variant of the Newton-Raphson method. The functional scheme of this important part of the program is shown in figure \ref{fig8}.  We wish to stress that although the number of variables can be quite large (more than $10^7$ loop variables), convergence is reasonably fast, because the initial concentration values are invariably very close to the final ones.

\subsection*{Cell motion}

Cell motion is also regulated by differential equations and the solution uses a strategy based on a semi-implicit method (described in detail in the supporting text). Volume growth is regulated by the part that handles metabolism and diffusion, therefore it is loosely coupled to cell motion. However we have implemented an updating mechanism that effectively decouples the two parts of the program: this means that the program can use multithreading with shared memory and exploit the features of multicore processors. 

\subsection*{Cellular events}

This part of the program handles discrete events, like cell-cycle transitions, mitosis and cell death. In case of mitosis it also initializes the daughter cells -- using the metabolic variables of the mother cell -- and allocates memory for the new cells. 

\subsection*{Geometry and topology of cell cluster}

Geometrical and topological informations are updated here, using calls to CGAL methods  \cite{r31}  that compute the convex hull of the cluster of cells, the Delaunay triangulation of cell centers, and the alpha shape of the cluster -- with an alpha parameter \cite{r31} equal to $(2r_0)^2$ where $r_0$ is the average cell radius. This part of the program uses this basic information to set all relevant geometrical and topological variables in the program.

\subsection*{Summary statistics and dump on file}

The last step in the loop computes several statistics and outputs them on summary files. It also writes periodically the whole configuration of cells on file for further processing. 

\subsection*{Program termination}

The program repeats the loop until one of the stop conditions is met: either all cells are dead, or the program executed the required number of steps. The supporting information text contains additional considerations on algorithmic  complexity and on measured performance (see figure SF6 and figure SF7).

Additional processing to extract useful informations from the simulation data is performed with several standard tools, like {\it Mathematica} \cite{r44}

\section*{Supporting Information}

{\bf Supporting text}: provides technical details on the structure of the simulation program and includes tables ST1 to ST5.\\
{\bf Figure SF1}: sketch of the cell phases included in the simulation program.\\
{\bf Figure SF2}: sketch of the metabolic network.\\
{\bf Figure SF3}: the geometry and topology of diffusion.\\
{\bf Figure SF4}: pictorial representation of the interaction force between two cells.\\
{\bf Figure SF5}: the geometry of mitosis.\\
{\bf Figure SF6}: CPU time needed to simulate 1 hour, vs. the number of cells in the spheroid.\\
{\bf Figure SF7}: total CPU time vs. the total number of cells.\\
{\bf Figures SF8-10}: oxygen concentration.\\
{\bf Figures SF11-14}: extracellular glucose concentration. \\
{\bf Figures SF15-17}: extracellular glutamine concentration.\\
{\bf Figure SF18}: lactate concentration .\\
{\bf Figures SF19-21}: velocity in the plane of the slice.\\
{\bf Movie S1}: development of the necrotic core. \\
{\bf Movie S2}: flow of extracellular glucose.\\
{\bf Movie S3}: map of projected cell velocities.\\

\section*{Acknowledgments}


\bibliography{refs}

\begin{thebibliography}{10}
\providecommand{\url}[1]{\texttt{#1}}
\providecommand{\urlprefix}{URL }
\expandafter\ifx\csname urlstyle\endcsname\relax
  \providecommand{\doi}[1]{doi:\discretionary{}{}{}#1}\else
  \providecommand{\doi}{doi:\discretionary{}{}{}\begingroup
  \urlstyle{rm}\Url}\fi
\providecommand{\bibAnnoteFile}[1]{%
  \IfFileExists{#1}{\begin{quotation}\noindent\textsc{Key:} #1\\
  \textsc{Annotation:}\ \input{#1}\end{quotation}}{}}
\providecommand{\bibAnnote}[2]{%
  \begin{quotation}\noindent\textsc{Key:} #1\\
  \textsc{Annotation:}\ #2\end{quotation}}
\providecommand{\eprint}[2][]{\url{#2}}

\bibitem{r3}
Sutherland RM (1988) Cell and environment interactions in tumor microregions:
  the multicell spheroid model.
\newblock Science 240: 177-84.
\bibAnnoteFile{r3}

\bibitem{r4}
Bjerkvig R (1992) {Spheroid Culture in Cancer Research}.
\newblock Boca Raton, Fla.: CRC Press.
\bibAnnoteFile{r4}

\bibitem{r5}
Mueller-Klieser W (1997) Three-dimensional cell cultures: from molecular
  mechanisms to clinical applications.
\newblock Am J Physiol 273: C1109-23.
\bibAnnoteFile{r5}

\bibitem{r6}
Mueller-Klieser W (2000) Tumor biology and experimental therapeutics.
\newblock Crit Rev Oncol Hematol 36: 123-39.
\bibAnnoteFile{r6}

\bibitem{Gott2006}
Gottfried E, Kunz-Schughart LA, Andreesen R, Kreutz M (2006) {Brave little
  world: spheroids as an in vitro model to study tumor-immune-cell
  interactions}.
\newblock Cell Cycle 5: 691-695.
\bibAnnoteFile{Gott2006}

\bibitem{Fried2007}
Friedrich J, Ebner R, Kunz-Schughart LA (2007) {Experimental anti-tumor therapy
  in 3-D: spheroids--old hat or new challenge?}
\newblock Int J Radiat Biol 83: 849-871.
\bibAnnoteFile{Fried2007}

\bibitem{LinChang2008}
Lin RZ, Chang HY (2008) Recent advances in three-dimensional multicellular
  spheroid culture for biomedical research.
\newblock Biotechnology Journal 3: 1172.
\bibAnnoteFile{LinChang2008}

\bibitem{Hirsch2010}
Hirschhaeuser F, Menne H, Dittfeld C, West J, Mueller-Klieser W, et~al. (2010)
  Multicellular tumor spheroids: an underestimated tool is catching up again.
\newblock J Biotechnol 148: 3-15.
\bibAnnoteFile{Hirsch2010}

\bibitem{r7}
Casciari JJ, Sotirchos SV, Sutherland RM (1992) {Mathematical modelling of
  microenvironment and growth in EMT6/Ro multicellular tumour spheroids}.
\newblock Cell Prolif 25: 1-22.
\bibAnnoteFile{r7}

\bibitem{r8}
Venkatasubramanian R, Henson MA, Forbes NS (2006) Incorporating energy
  metabolism into a growth model of multicellular tumor spheroids.
\newblock J Theor Biol 242: 440-53.
\bibAnnoteFile{r8}

\bibitem{r9}
Sander LM, Deisboeck TS (2002) Growth patterns of microscopic brain tumors.
\newblock Phys Rev E Stat Nonlin Soft Matter Phys 66: 051901.
\bibAnnoteFile{r9}

\bibitem{r10}
Stein AM, Demuth T, Mobley D, Berens M, Sander LM (2007) A mathematical model
  of glioblastoma tumor spheroid invasion in a three-dimensional in vitro
  experiment.
\newblock Biophys J 92: 356-65.
\bibAnnoteFile{r10}

\bibitem{r11}
Jiang Y, Pjesivac-Grbovic J, Cantrell C, Freyer JP (2005) A multiscale model
  for avascular tumor growth.
\newblock Biophys J 89: 3884-94.
\bibAnnoteFile{r11}

\bibitem{r12}
Schaller G, Meyer-Hermann M (2005) {Multicellular tumor spheroid in an
  off-lattice Voronoi-Delaunay cell model}.
\newblock Phys Rev E Stat Nonlin Soft Matter Phys 71: 051910.
\bibAnnoteFile{r12}

\bibitem{r13}
Kim Y, Stolarska MA, Othmer HG (2007) {A Hybrid Model For Tumor Spheroid Growth
  In Vitro I: Theoretical Development And Early Results}.
\newblock Mathematical Models and Methods in Applied Sciences 17, Suppl.:
  1773--1798.
\bibAnnoteFile{r13}

\bibitem{r14}
Engelberg JA, Ropella GEP, Hunt CA (2008) Essential operating principles for
  tumor spheroid growth.
\newblock BMC Syst Biol 2: 110.
\bibAnnoteFile{r14}

\bibitem{r15}
Nederman T, Norling B, Glimelius B, Carlsson J, Brunk U (1984) Demonstration of
  an extracellular matrix in multicellular tumor spheroids.
\newblock Cancer Res 44: 3090-7.
\bibAnnoteFile{r15}

\bibitem{r16}
Dorie MJ, Kallman RF, Rapacchietta DF, Van~Antwerp D, Huang YR (1982) Migration
  and internalization of cells and polystyrene microsphere in tumor cell
  spheroids.
\newblock Exp Cell Res 141: 201-9.
\bibAnnoteFile{r16}

\bibitem{r17}
Chignola R, Foroni R, Franceschi A, Pasti M, Candiani C, et~al. (1995)
  Heterogeneous response of individual multicellular tumour spheroids to
  immunotoxins and ricin toxin.
\newblock Br J Cancer 72: 607-14.
\bibAnnoteFile{r17}

\bibitem{r18}
Bortner CD, Cidlowski JA (2002) Apoptotic volume decrease and the incredible
  shrinking cell.
\newblock Cell Death Differ 9: 1307-10.
\bibAnnoteFile{r18}

\bibitem{r19}
Chignola R, Milotti E (2005) A phenomenological approach to the simulation of
  metabolism and proliferation dynamics of large tumour cell populations.
\newblock Phys Biol 2: 8-22.
\bibAnnoteFile{r19}

\bibitem{r20}
Chignola R, {Del Fabbro} A, {Dalla Pellegrina} C, Milotti E (2007) Ab initio
  phenomenological simulation of the growth of large tumor cell populations.
\newblock Phys Biol 4: 114-33.
\bibAnnoteFile{r20}

\bibitem{r21}
Chignola R, Milotti E (2004) Numerical simulation of tumor spheroid dynamics.
\newblock Physica A: Statistical Mechanics and its Applications 338: 261 - 266.
\bibAnnoteFile{r21}

\bibitem{r22}
Alberts B, Wilson JH, Hunt T (2008) {Molecular Biology of the Cell}.
\newblock New York: Garland Science, 5th edition.
\bibAnnoteFile{r22}

\bibitem{r23}
Chignola R, {Dalla Pellegrina} C, {Del Fabbro} A, Milotti E (2006) Thresholds,
  long delays and stability from generalized allosteric effect in protein
  networks.
\newblock Physica A: Statistical and Theoretical Physics 371: 463 - 472.
\bibAnnoteFile{r23}

\bibitem{r24}
Milotti E, {Del Fabbro} A, {Dalla Pellegrina} C, Chignola R (2007) Dynamics of
  allosteric action in multisite protein modification.
\newblock Physica A: Statistical Mechanics and its Applications 379: 133 - 150.
\bibAnnoteFile{r24}

\bibitem{r25}
Chignola R, Schenetti A, Andrighetto G, Chiesa E, Foroni R, et~al. (2000)
  Forecasting the growth of multicell tumour spheroids: implications for the
  dynamic growth of solid tumours.
\newblock Cell Prolif 33: 219-29.
\bibAnnoteFile{r25}

\bibitem{r26}
Yu P, Mustata M, Turek JJ, French PMW, Melloch MR, et~al. (2003) Holographic
  optical coherence imaging of tumor spheroids.
\newblock Appl Phys Lett 83: 575--577.
\bibAnnoteFile{r26}

\bibitem{r27}
Milotti E, {Del Fabbro} A, Chignola R (2009) Numerical integration methods for
  large-scale biophysical simulations.
\newblock Computer Physics Communications 180: 2166 - 2174.
\bibAnnoteFile{r27}

\bibitem{r28}
Cornish-Bowden A (1979) Fundamentals of enzyme kinetics.
\newblock London: Butterworths.
\bibAnnoteFile{r28}

\bibitem{r29}
O'Rourke J (1998) {Computational geometry in C}.
\newblock Cambridge, UK,: Cambridge University Press, 2nd ed edition.
\bibAnnoteFile{r29}

\bibitem{r30}
de~Berg M, Cheon O, van Kreveld M, Overmars M (2008) {Computational Geometry:
  Algorithms and Applications}.
\newblock Berlin: Springer, 3rd ed edition.
\bibAnnoteFile{r30}

\bibitem{r31}
\textsc{Cgal}, {C}omputational {G}eometry {A}lgorithms {L}ibrary.
\newblock Http://www.cgal.org.
\bibAnnoteFile{r31}

\bibitem{r32}
Guibas LJ, Knuth DE, Sharir M (1992) {Randomized incremental construction of
  Delaunay and Voronoi diagrams}.
\newblock Algorithmica 7: 381-413.
\bibAnnoteFile{r32}

\bibitem{r33}
Dwyer RA (1991) {Higher-dimensional Voronoi diagrams in linear expected time}.
\newblock Discrete and Computational Geometry 6: 343-367.
\bibAnnoteFile{r33}

\bibitem{r36}
Flekk{\o}y EG, Coveney PV, Fabritiis GD (2000) Foundations of dissipative
  particle dynamics.
\newblock Phys Rev E 62: 2140--2157.
\bibAnnoteFile{r36}

\bibitem{r37}
Walenta S, Doetsch J, Mueller-Klieser W, Kunz-Schughart LA (2000) Metabolic
  imaging in multicellular spheroids of oncogene-transfected fibroblasts.
\newblock J Histochem Cytochem 48: 509-22.
\bibAnnoteFile{r37}

\bibitem{r38}
Mueller-Klieser W, Freyer JP, Sutherland RM (1986) Influence of glucose and
  oxygen supply conditions on the oxygenation of multicellular spheroids.
\newblock Br J Cancer 53: 345-53.
\bibAnnoteFile{r38}

\bibitem{r39}
Alvarez-P{\'e}rez J, Ballesteros P, Cerd{\'a}n S (2005) {Microscopic images of
  intraspheroidal pH by 1H magnetic resonance chemical shift imaging of pH
  sensitive indicators}.
\newblock MAGMA 18: 293-301.
\bibAnnoteFile{r39}

\bibitem{r40}
Acker H, Carlsson J, Holtermann G, Nederman T, Nylen T (1987) {Influence of
  Glucose and Buffer Capacity in the Culture Medium on Growth and pH in
  Spheroids of Human Thyroid Carcinoma and Human Glioma Origin}.
\newblock Cancer Res 47: 3504-3508.
\bibAnnoteFile{r40}

\bibitem{r41}
Sutherland RM, Sordat B, Bamat J, Gabbert H, Bourrat B, et~al. (1986)
  {Oxygenation and Differentiation in Multicellular Spheroids of Human Colon
  Carcinoma}.
\newblock Cancer Res 46: 5320-5329.
\bibAnnoteFile{r41}

\bibitem{r42}
Khaitan D, Chandna S, Arya MB, Dwarakanath BS (2006) Establishment and
  characterization of multicellular spheroids from a human glioma cell line;
  implications for tumor therapy.
\newblock J Transl Med 4: 12.
\bibAnnoteFile{r42}

\bibitem{Deis}
Deisboeck TS, Couzin ID (2009) Collective behavior in cancer cell populations.
\newblock Bioessays 31: 190-7.
\bibAnnoteFile{Deis}

\bibitem{r44}
Wolfram~Research I (2008) Mathematica.
\newblock Champaign, Illinois: Wolfram Research, Inc., version 7.0 edition.
\bibAnnoteFile{r44}

\bibitem{r43}
Yuhas JM, Li AP, Martinez AO, Ladman AJ (1977) A simplified method for
  production and growth of multicellular tumor spheroids.
\newblock Cancer Res 37: 3639-43.
\bibAnnoteFile{r43}

\end{thebibliography}

\clearpage

\section*{Figures}

\clearpage

\section*{Tables}

\begin{table}[htdp]
\caption{
\bf{Comparisons with experimental parameters}}
{
\begin{tabular}{|c|c|c|c|c|}
\hline
{\bf Parameter} & {\bf Simulation} & {\bf Experiments} & $^7${\bf Cell type} & {\bf Ref.} \\
\hline 
$^1$Glucose uptake (kg s$^{-1}$ m$^{-3}$) & $1.44 \cdot 10^{-3}$ & $5.4-12.6 \cdot 10^{-3}$ & Rat-T1, MR1 & \cite{r37} \\
\hline 
$^1$Lactate release (kg s$^{-1}$ m$^{-3}$) & $1.35 \cdot 10^{-3}$ & $5.4-9 \cdot 10^{-3}$ & Rat-T1, MR1 & \cite{r37} \\
\hline
$^2$pO$_2$ (mmHg) & 7 & 0-20 & Rat-T1 & \cite{r37}\\
 &  &  0-40  &  MR1 & \cite{r37} \\
& & 20-60 & EMT6/Ro & \cite{r38} \\
\hline
$^3$pH & 6.7 & 6.6 & C6, H35 &\cite{r39} \\
 &  &  6.96-6.99 & U118-MG, HTh7 &  \cite{r40} \\
\hline
$^4$$\Delta$pH & 0.77 & 0.41 & U118-MG & \cite{r40} \\
 &  & 0.49 $\pm$ 0.08 & HTh7 & \cite{r40} \\
\hline
$^5$Viable cell rim thickness ($\mu$m) & 155 & 200 & EMT6/Ro & \cite{r38} \\
& & 142 $\pm$ 16 & HTh7 & \cite{r40}\\
& & 310 $\pm$ 28 & U118-MG & \cite{r40}\\
& & 198 $\pm$ 27 & Col12 & \cite{r41}\\
& & 225 $\pm$ 26 & HT29 & \cite{r41}\\
\hline
$^6$Hypoxic rim thickness ($\mu$m) & 98 & 44 $\pm$ 52 &Col12 & \cite{r41} \\
&  & 44 $\pm$ 52 & HT29  & \cite{r41}  \\
\hline
Cell cycle distribution (\%) & G0/G1 = 57.3 & G0/G1 = 58 $\pm$ 4 & BMG-1 & \cite{r42} \\
& S = 21.6 &  S = 19 $\pm$ 1 &   &\\
& G2/M = 21.1 & G2/M = 23 $\pm$ 1 &  & \\
\hline
\end{tabular} 
}
\begin{flushleft}
{Metabolic and histologic parameters in spheroids of  approximately 500 $\mu$m diameter: comparison between a single, large simulation, carried out with the parameters listed in tables ST4 and ST5 in the supporting text, and experimental data.}
\end{flushleft}
\label{tab1}
\end{table}
Notes:
\begin{enumerate}
\item Rate of glucose uptake or lactate release per viable spheroid volume (see \cite{r37}).
\item Central pO$_2$ tension (experiments) or estimated in the centroid (simulations).
\item pH has been determined in the central region of the spheroids. This corresponds to a sphere of radius $\approx 100 \mu$m about the centroid of the spheroid.
\item Difference between environmental pH and pH 200 $\mu$m below the spheroid surface.
\item In our simulations the viable cell rim thickness corresponds to the distance between the spheroid surface and the inner shell where only 5\% of the cells are still alive. Experimental values have been determined by histology.
\item These values corresponds to the radius of the necrotic core.
\item Cell types are as follows: Rat-T1 = T24Ha-ras-transfected Rat1 cells (Rat1 = spontaneously immortalized rat embryo fibroblasts); MR1 = myc/ T24Ha-ras-cotransfected rat embryo fibroblasts; EMT6/Ro = mouse mammary tumor cells; C6 = rat glioma cells; H35 = rat hepatoma cells; U118-MG = human glioblastoma cells; HTh7 = human tyroid carcinoma cells; Col12 = moderately differentiated human colon adenocarcinoma ; HT29 = poorly differentiated human colon adenocarcinoma; BMG-1 = human glioma cells.
\end{enumerate}

\end{document}